\documentclass[twocolumn,twoside,slac_two]{revtex4}
\usepackage{graphicx}
\usepackage{fancyhdr}
 \newcommand{\beq}{\begin{equation}}
 \newcommand{\eeq}{\end{equation}}
 \newcommand{\bqa}{\begin{eqnarray}}
 \newcommand{\eqa}{\end{eqnarray}}

\def\ket#1{\left| #1\right\rangle}

\newcommand{\Kz}{\mbox{$K^{0}$}}
\newcommand{\Kzbar}{\mbox{$\overline{K^{0}}$}}

\newcommand{\KL}{\mbox{$K_{L}$}}
\newcommand{\KS}{\mbox{$K_{S}$}}

\newcommand{\tauS}{\mbox{$\tau_{S}$}}

\newcommand{\delm}{\mbox{$\Delta m$}}

\def\epe{\epsilon'\!/\epsilon}

\newcommand{\epsrat}{\mbox{$\epsilon^{\prime}\!/\epsilon$}}
\newcommand{\reepoe}{\mbox{$Re(\epsrat)$}}
\newcommand{\imepoe}{\mbox{$Im(\epsrat)$}}
\newcommand{\ktev}{\mbox{KTeV}}
\newcommand{\etapm}{\mbox{$\eta_{+-}$}}
\newcommand{\etazz}{\mbox{$\eta_{00}$}}
\newcommand{\phipm}{\mbox{$\phi_{+-}$}}
\newcommand{\phizz}{\mbox{$\phi_{00}$}}
\newcommand{\phisw}{\mbox{$\phi_{SW}$}}
\newcommand{\phiep}{\mbox{$\phi_{\epsilon}$}}
\newcommand{\delphi}{\mbox{$\Delta \phi$}}

\def\kchrg{K\to\pi^+\pi^-}
\def\kneut{K\to\pi^0\pi^0}

\def\kpi0{K_{L}\to 3\pi^0}
\def\ke3{K_{L}\to\pi^{\pm}e^{\mp}\nu}
\def\km3{K_{L}\to\pi^{\pm}\mu^{\mp}\nu}
\def\k2pi{K_{L} \to \pi^+\pi^-}

\newcommand{\Kpp}{\mbox{$K\rightarrow\pi\pi$}}
\newcommand{\Kpm}{\mbox{$K\rightarrow\pi^{+}\pi^{-}$}}
\newcommand{\Kzz}{\mbox{$K\rightarrow\pi^{0}\pi^{0}$}}

\newcommand{\Kethree}{\mbox{$K_{e3}$}}

\newcommand{\KLpp}{\mbox{$K_{L}\rightarrow\pi\pi$}}
\newcommand{\KSpp}{\mbox{$K_{S}\rightarrow\pi\pi$}}
\newcommand{\KLpm}{\mbox{$K_{L}\rightarrow\pi^{+}\pi^{-}$}}
\newcommand{\KSpm}{\mbox{$K_{S}\rightarrow\pi^{+}\pi^{-}$}}
\newcommand{\KLzz}{\mbox{$K_{L}\rightarrow\pi^{0}\pi^{0}$}}
\newcommand{\KSzz}{\mbox{$K_{S}\rightarrow\pi^{0}\pi^{0}$}}

\newcommand{\Kzzz}{\mbox{$K_L \rightarrow \pi^{0}\pi^{0}\pi^{0}$}}

\newcommand{\ppc}{\mbox{$\pi^{+}\pi^{-}$}}

\newcommand{\ppn}{\mbox{$\pi^{0}\pi^{0}$}}

\newcommand{\pzpz}{\pi^{0}\pi^{0}}

\newcommand{\keth}{\mbox{$\pi^{\pm} e^{ \mp}\nu_e$}}

\newcommand{\zzz}{\mbox{$\pi^{0}\pi^{0}\pi^{0}$}}

\newcommand{\eu}{ \times 10^{-4}}

\newcommand{\degs}{^{\circ}}
\newcommand{\delmunits}{\mbox{$\times 10^{6}~\hbar {\rm s}^{-1}$}}
\newcommand{\tausunits}{\mbox{$\times 10^{-12}~{\rm s}$}}

\def\dmswval{5269.9}
\def\tsswval{89.623}

\def\dmswerr{  12.3}
\def\tsswerr{ 0.047}

\def\dphcptval{ 0.30}

\def\dphcpterr{ 0.35}

\def\phpmval{ 43.76}
\def\phpmerr{  0.64}
\def\phzzval{ 44.06}
\def\phzzerr{  0.68}

\def\dphswval{  0.40}
\def\dphswerr{  0.56}

\begin{document}

\title{The Final Measurement of $\epe$ from KTeV}

%

\author{E. Worcester}
\affiliation{University of Chicago, Chicago, Illinois}

\begin{abstract}
We present precise measurements of CP and CPT symmetry
based on the full dataset of $\Kpp$ decays collected by the KTeV experiment 
at Fermi National Accelerator Laboratory during 1996, 1997, and 1999.
This dataset contains about 15 million $\kneut$ and 70 million $\kchrg$ decays.  
We measure the direct CP violation parameter$\reepoe$ = (19.2 $\pm 2.1)\eu$.
We find the $K_L$-$K_S$ mass difference $\delm$~=~(5265 $\pm$ 10)$\delmunits$
and the $K_S$ lifetime $\tauS$ = (89.62 $\pm$ 0.05)$\tausunits$.  We test CPT
symmetry by finding the phase of the indirect CP violation parameter 
$\epsilon$, $\phiep$~=~(44.09 $\pm$ 1.00)$\degs$, and the difference of the relative 
phases between the CP violating and CP conserving decay amplitudes for 
$\Kpm$ ($\phipm$) and for $\Kzz$ ($\phizz$), $\delphi$ = (0.29 $\pm$ 0.31)$\degs$.
These results are consistent with other experimental results and
with CPT symmetry.

\end{abstract}

\maketitle

\thispagestyle{fancy}


\section{Introduction}
Violation of CP symmetry occurs in the neutral kaon system in
two different ways.  The dominant effect is the
result of an asymmetry in the mixing of $\Kz$ and $\Kzbar$ such
that $\KL$ and $\KS$ are not CP eigenstates.  This effect is
parameterized by $\epsilon$ and is called indirect CP violation.
The other effect, called direct CP violation, occurs in the
$\Kpp$ decay process and is parameterized by $\epsilon'$.
Direct CP violation affects the decay rates of $\Kpm$ and $\Kzz$
differently, so it is possible to measure the level of direct
CP violation by comparing $\etapm$ and $\etazz$:
\begin{equation}
 \begin{array}{ccccc}
\etapm & = &\frac{A(\KLpm)}{A(\KSpm)}\ & = & \epsilon + \epsilon'
\\
\\
\etazz & = &\frac{A(\KLzz)}{A(\KSzz)}\ & = & \epsilon - 2\epsilon'
\\
\\
\reepoe & \approx & \frac{1}{6}\ ( |\frac{\etapm}{\etazz}\ |^2 - 1).& &
\\
 \end{array}
\label{eq:reepoe}
\end{equation}
Measurements of $\pi\pi$ phase shifts~\cite{ochs} show that,
in the absence of CPT violation, the phase of $\epsilon'$ is
approximately equal to that of $\epsilon$. Therefore, $\reepoe $ is a
measure of direct CP violation and $\imepoe $ is a measure of CPT
violation.

\begin{figure}
\centering
\includegraphics[width=80mm]{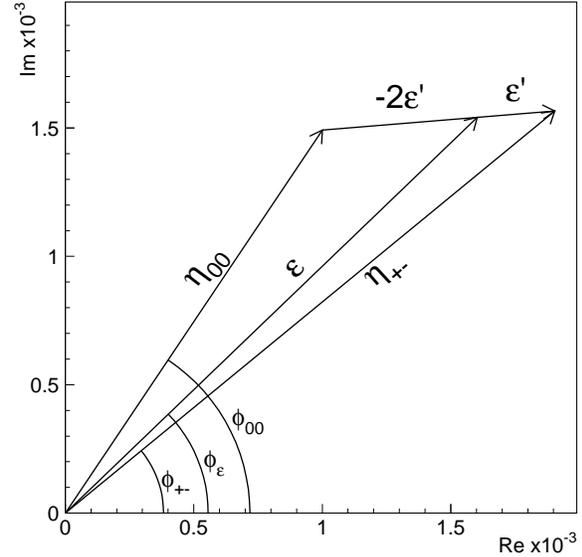}
\caption{Diagram of CP violating kaon parameters.  For this illustration, the
parameters have the central values measured by KTeV and
the value of $\epsilon'$ is scaled by a factor of 50. Although they
appear distinct in this diagram,
note that $\phipm$ and 
$\phizz$ are consistent with each other within experimental errors.}
\label{fig:kaonparams}
\end{figure}

For small $|\epsilon'/\epsilon|$, $\imepoe$ is related
to the phases of $\etapm$ and $\etazz$ by
\begin{equation}
\begin{array}{lcl}
  \phipm &\approx &\phiep  + \imepoe  \\
  \phizz &\approx &\phiep  - 2 \imepoe  \\
  \delphi &\equiv &\phizz  - \phipm   \approx -3  \imepoe~.
\end{array}
  \label{eq:delphimpe}
\end{equation}
The relation of the complex parameters $\etapm$, $\etazz$,
$\epsilon$, and $\epsilon'$ is illustrated in Fig.~\ref{fig:kaonparams} using
the central values measured by the KTeV experiment.

Experimental results have established that $\reepoe$ is non-zero
\cite{prl:731,pl:na31,prl:pss,na48:reepoe}.  In 2003, KTeV
reported $\reepoe = (20.7 \pm 2.8)\eu$ based on data from 1996 and 
1997\cite{prd03}.
We now report the 
final measurement of $\reepoe$ from KTeV.
The measurement is based on 85 million reconstructed \Kpp\ decays
collected in 1996 1997, and 1999.
This full sample is two 
times larger than, and contains, the sample on which the previous
results are based.
We also present measurements of
the kaon parameters $\delm$ and $\tauS$,
and tests of CPT symmetry based on measurements of
$\delphi$ and $\phipm -\phisw$.

For these results we have made significant improvements to the data 
analysis and the Monte Carlo simulation.
The full dataset, including those data used in the previous analysis,
has been reanalyzed using the improved reconstruction and simulation.
These results supersede the previously published results
from KTeV.
In this presentation, we will focus primarily on improvements to the
$\kneut$ analysis which have reduced the systematic uncertainty
in $\reepoe$
relative to the previous KTeV result.

\section{The KTeV Experiment}
The measurement of \reepoe\ requires a source of $K_L$ and $K_S$
decays, and a detector to reconstruct the charged ($\ppc$)
and neutral ($\ppn$) final states.
The strategy of the \ktev\ experiment is to produce two identical
$K_L$ beams, and then to pass one of the beams through a
``regenerator.''
The beam that passes through the regenerator is called the
regenerator beam,
and the other beam is called the vacuum beam.
The regenerator creates a coherent
$\ket{K_L}+\rho\ket{K_S}$ state,
where $\rho$, the regeneration amplitude, is a physical property
of the regenerator.  The regenerator is designed
such that most of the \Kpp\ decays
downstream of the regenerator are from the $K_S$ component.
The charged spectrometer is the 
primary detector for reconstructing $\Kpm$ decays and the 
pure Cesium Iodide (CsI) calorimeter 
is used to reconstruct the four photons from $\Kzz$ decays.  
A Monte Carlo simulation is used to correct for the acceptance
difference between \Kpp\ decays in the two beams,
which results from the very different $K_L$ and $K_S$ lifetimes.
The measured quantities are the vacuum-to-regenerator
``single ratios''  for \Kpm\ and \Kzz\ decay rates.
These single ratios are proportional to
$|\etapm/\rho|^2$ and $|\etazz/\rho|^2$,
and the ratio of these two quantities gives $\reepoe$
via Eq.~\ref{eq:reepoe}.

\subsection{The KTeV Detector}
\begin{figure}
\begin{centering}
\includegraphics[width=80mm]{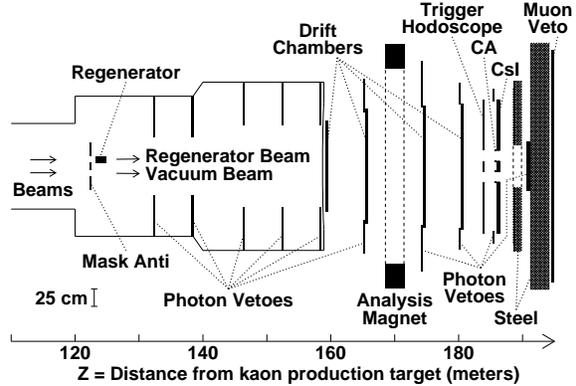}
\caption{The KTeV Detector}
\label{fig:det}
\end{centering}
\end{figure}

The KTeV detector (Fig. \ref{fig:det}) consists of a charged spectrometer 
to reconstruct
$\Kpm$ decays, a pure CsI electromagnetic calorimeter to reconstruct
$\Kzz$ decays, a veto system to reduce background, and a three-level trigger to
select events.  Two virtually identical neutral kaon beams are incident
on the detector; a movable active regenerator is placed in
one of these beams to provide a coherent mixture of $\KL$ and $\KS$.
In this manner, we collect $\KLpp$ and $\KSpp$ decays simultaneously
so that many systematic effects cancel in the ratios used to calculate
\reepoe.  

The KTeV spectrometer consists of four square drift chambers and a
large dipole magnet.  Each drift chamber measures charged-particle
positions in two planes each in the $x$ and $y$ views.  The drift
chamber planes have a hexagonal cell geometry formed by six field-shaping
wires surrounding one sense wire, and the two planes in each view are
offset to resolve position ambiguities. There are a total of 1972 sense
wires in the four drift chambers.  The magnet imparts a kick of 412 MeV/c
in the horizontal plane.
The spectrometer measures the momenta of charged particles with
an average resolution of ~0.4\%.  The $\Kpm$ reconstruction achieves a 
$z$-vertex 
resolution of 5-30 cm and a mass resolution of 1.5 MeV/$c^{2}$.

\begin{figure}
\begin{center}
\includegraphics[width=80mm]{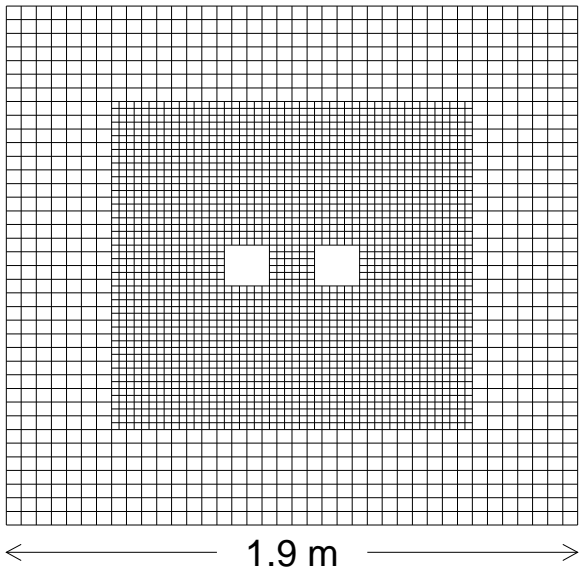}
\caption{Beamline view of the KTeV CsI calorimeter,
        showing the 868 larger outer
        crystals and the 2232 smaller inner crystals.
        Each beam hole size is  $15\times 15$~cm$^2$ and
        the two beam hole centers are separated by 0.3~m.}
\label{fig:csilayout}
\end{center}
\end{figure}

The CsI calorimeter measures the energies and positions of photons from
the electromagnetic decay of the neutral pions in $\Kzz$ decays.  
It consists of 3100 pure CsI crystals viewed by photomultiplier tubes.
The layout of the $1.9 \times 1.9$~m$^2$ calorimeter is shown in
Fig.~\ref{fig:csilayout}.
There are 2232 $2.5 \times 2.5$ cm$^2$ crystals in the central region,
and 868 $5 \times 5$ cm$^2$ crystals surrounding the smaller crystals.
The crystals are all 50~cm (27 radiation lengths) long.   
Momentum analyzed electrons and positrons from 
$\ke3$ decays ($\Kethree$) are used to calibrate the CsI energy scale 
to 0.02\%.
The CsI calorimeter has
an average energy resolution of 0.6\%.  The reconstructed decay vertex 
of the neutral
pion is directly related to the energies and positions of the photons:
\begin{equation}
\label{eq:zcalc}
Z_{\pi^0} = Z_{CsI} - \frac{r_{12}\sqrt{E_{1}E_{2}}}{m_{\pi^0}},\
\end{equation}
where $r_{12}$ is the transverse distance between the photons at the CsI
and $E_1$ and $E_2$ are the photon energies.
The $\Kzz$ reconstruction achieves a $z$-vertex resolution of 20-30 cm
and a mass resolution of 1.5 MeV/$c^{2}$.

KTeV uses an extensive veto system to reject events coming from
interactions in the regenerator and to reduce background from kaon
decays into non-$\pi\pi$ final states.  The veto system consists of
a number of lead-scintillator detectors surrounding the decay region
and the primary detectors.  A three-level trigger, 
consisting of fast detector signals at Level 1, processing by custom
electronics at Level 2, and a software filter at Level 3, is
used to select events.

\subsection{Monte Carlo Simulation}
KTeV uses a Monte Carlo (MC) simulation to calculate the detector
acceptance and to model background to the signal modes.  
The very different $\KL$ and $\KS$ lifetimes lead to very different 
$z$-vertex distributions in the 
vacuum and regenerator beams.  We determine the 
detector acceptance as a function of kaon decay vertex and energy including 
the effects of geometry, detector response, and resolutions.  
The simulation of the detector geometry is based both on data and survey
measurements.  Many aspects of the tracing and detector response are based on
libraries created by GEANT\cite{geant} simulations.
To help verify the accuracy of the MC simulation, we collect 
and study decay modes with approximately ten times higher statistics than 
the $\Kpp$ signal samples, such as $\ke3$ and \Kzzz.

Many improvements have been made to the MC simulation since 
the previous result was published in 2003.  We have improved
the simulation of electromagnetic showers
to include the effects of incident particle angles
and to simulate the effects of wrapping and shims in
the CsI calorimeter.  The GEANT library used for the previous analysis
was binned in energy and incident position; the effect of angles was
approximated by shifting the incident position based on the angle of 
incidence.  The shower library has now been expanded to include nine angles 
(-35 mrad to 35 mrad) for photons and 15 angles (-85 mrad to 85 mrad) for 
electrons.  Electrons angles may be larger than photon angles because of
the momentum kick imparted by the analyzing magnet.  Differences between 
the library angle and the desired angle are 
approximated by shifting the incident position.
The particle energy cutoff applied 
in the GEANT shower library generation
has been lowered from 600 keV to 50 keV for electrons; the photon
cutoff of 50 keV is unchanged.
Sixteen showers per bin have been generated.  
Energy deposits are corrected for energy lost in the 12 $\mu$m mylar
wrapping around the CsI crystals and the shims that are present
between some rows of CsI crystals.

\begin{figure}
\begin{center}
\includegraphics[width=80mm]{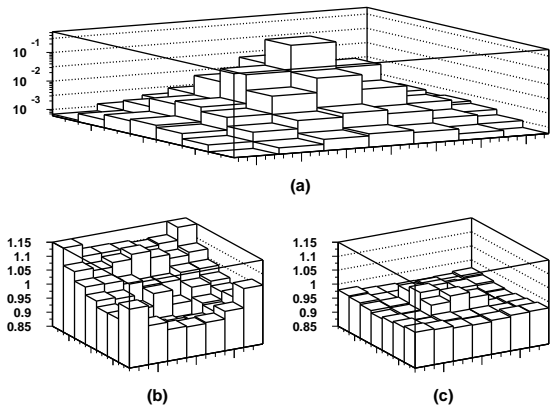}
\caption{Data-MC comparison of fraction of energy in each of the 49 CsI
crystals in an electron shower.  
(a) The fraction of energy in each of the 49 CsI crystals in an electron
shower for data.  (b) 2003 data/MC ratio (c) current data/MC ratio}
\label{fig:ke3shwr2}
\end{center}
\end{figure}

The current Monte Carlo produces a significantly better 
simulation of electromagnetic showers in the CsI.  
Figure \ref{fig:ke3shwr2}
shows a data-MC comparison of the fraction of energy 
in each of the 49 CsI crystals in a
shower relative to the total reconstructed shower energy for 
electrons from $\ke3$ decays.  The majority of the energy is deposited in the
central crystal since the Moliere radius of CsI is 3.8 cm.  
These particular plots are made for 16-32 GeV electrons with incident angles 
of 20-30 mrad, but the quality of 
agreement is similar for other energies and angles.  The data-MC disagreement 
improves from up to 15\% for
the 2003 MC to less than 5\% for the current MC.  This improvement in the
modeling of electromagnetic shower shapes leads to important reductions 
in the systematic
uncertainties associated with the reconstruction of photon showers from
$\kneut$ decays.

We have improved the tracing of charged particles through the detector
with more complete treatments of ionization energy loss, Bremsstrahlung,
delta rays, and hadronic interactions in the drift chambers.  
The position resolution of the drift chambers was previously treated
as flat across the cell; the dependence of the resolution upon position
within the cell is now included in the simulation.
We have also have updated a number of parameters that go into the 
kaon propogation and decay calculations.

\section{Data Analysis}
\begin{figure}
\begin{center}
\includegraphics[width=80mm]{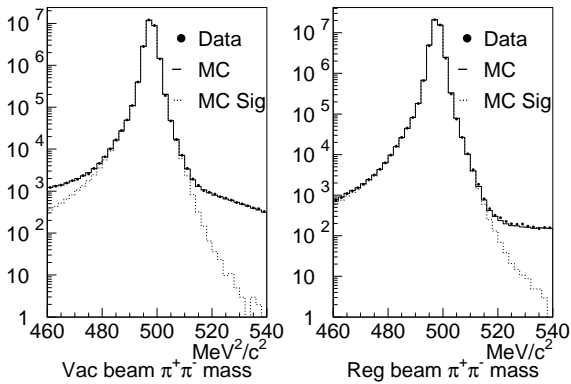}
\end{center}
\caption{\label{fig:massch}
$\pi^+\pi^-$ invariant mass distribution for $\kchrg$ candidate
events. The data distribution is shown as dots, the $\kchrg(\gamma)$ 
signal MC (MC Sig) is shown as a dotted histogram and the sum
of signal and background MC is shown as a solid histogram.}
\end{figure}

The $\kchrg$ analysis consists primarily of the reconstruction of tracks 
in the spectrometer.  The vertices and momenta of the tracks are used to 
calculate kinematic quantities describing the decay.  The $\kchrg$
invariant mass distributions for each beam are shown in Fig. 
\ref{fig:massch}.

To reconstruct 
$\kneut$ decays, we measure the energies and 
positions of each cluster of energy in the CsI.  
A number of 
corrections are then made to the measured particle energies based on our 
knowledge of the CsI performance and the reconstruction algorithm.  The
precision of the CsI energy and position reconstruction is crucial to the
$\kneut$ analysis and has been improved significantly since the previous
publication.  

We determine the energy deposit in each block of the CsI by converting
the digitized information to energies using constants for each channel 
that are determined from the electron calibration.  The laser correction,
which is measured using an in-situ laser and corrects for spill-to-spill 
drifts in each channel's gain, is applied to each block energy.
We define ``clusters,'' which are 7$\times$7 arrays of small blocks or 
3$\times$3 arrays of large blocks, centered on a ``seed block,'' which 
contains a local energy maximum.  
The cluster energy must be corrected for a number of geometric and 
detector effects.  We apply ``block-level'' corrections, which adjust 
the energy in each block that makes up the cluster, and
``cluster-level'' corrections, which are multiplicative corrections to the
total cluster energy.

\begin{figure}
\begin{center}
\includegraphics[width=80mm]{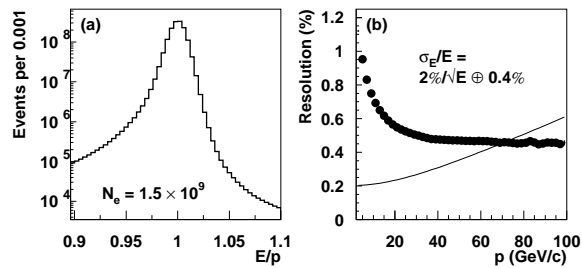}
\caption{$\Kethree$ electrons after all corrections. (a) E/p for 1.5 
$\times$ 10$^9$ electrons. (b) Energy resolution.  The fine
curve shows the momentum resolution function that has been subtracted from 
the E/p resolution to
find the energy resolution.}
\label{fig:resfinal}
\end{center}
\end{figure}

The quality of the calibration and the CsI performance is evaluated by 
analzying electrons from the calibration sample with all 
corrections applied.  The electron calibration for 1996, 1997, and 1999 is 
based on 1.5 billion total electrons.  Figure \ref{fig:resfinal} shows the 
E/p distribution and the energy resolution as a function of momentum after 
all corrections.
The final energy resolution of the calorimeter is 
$\sigma_E/E \simeq 2\% / \sqrt{E} \oplus 0.4\%$, where E is in GeV.

The position of a cluster is reconstructed by calculating the fraction of 
energy in neighboring columns and rows of the cluster.  The position algorithm 
uses a map that is based on the uniform photon illumination across each 
crystal to convert 
these ratios to a position within the seed block.  

\begin{figure}
\begin{center}
\includegraphics[width=80mm]{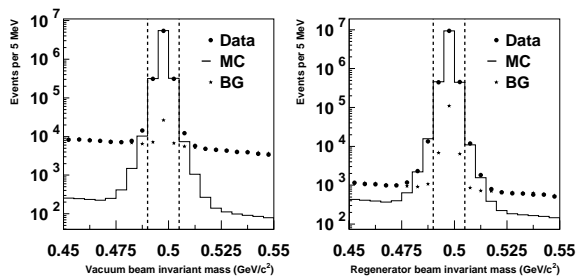}
\caption{$\kneut$ m$_{\pzpz}$ distributions for for data (dots) and signal 
MC (histogram) in the vacuum (left) and regenerator (right) beams.  The sum 
of the background MC is also shown (stars). The dashed lines indicate our 
cuts.}
\label{fig:mkcut}
\end{center}
\end{figure}

We use the cluster energies and positions along with the known pion mass 
to determine which pair of photons is associated with which neutral pion 
from the kaon decay and to calculate the decay vertex, the center of
energy, and the $\pzpz$ invariant mass.  
The $\kneut$ invariant mass distributions
for each beam are shown in Fig. \ref{fig:mkcut}.

\begin{figure}
\begin{center}
\includegraphics[width=80mm]{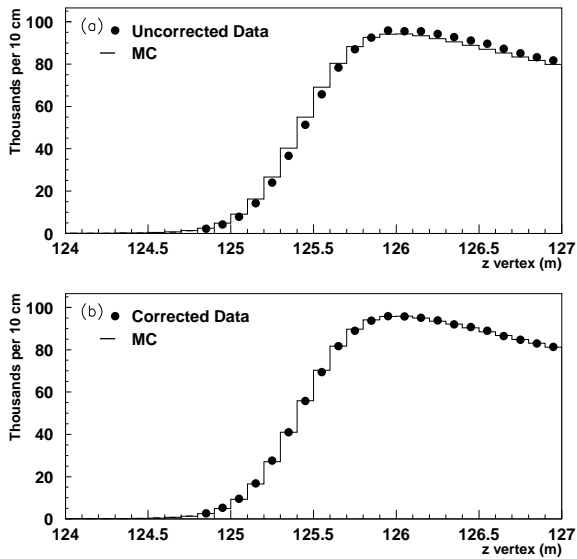}
\caption{Regenerator beam $\kneut$ $z$-vertex distribution near the
regenerator for 1999 data and Monte Carlo.  (a) Uncorrected data.  (b) Data
with energy scale correction applied.}
\label{fig:edgematch}
\end{center}
\end{figure}

\begin{figure}
\begin{center}
\includegraphics[width=80mm]{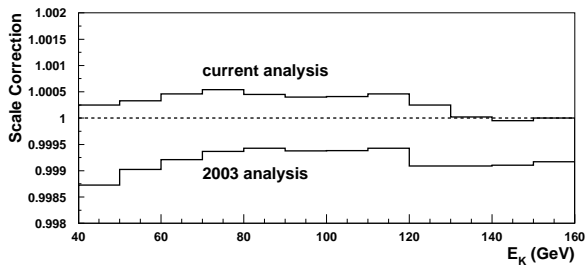}
\caption{Change in the final energy scale adjustment 
relative to the 2003 analysis.  The dashed line represents no data-MC
mismatch.}
\label{fig:scalechange}
\end{center}
\end{figure}

\looseness+1
For $\Kzz$ decays, the $z$ vertex is determined using only the positions
and energies of the four photons in the final state.  Therefore, the 
measured $z$ vertex is dependent upon the absolute energy scale of the 
CsI calorimeter.  The energy scale is set using
electrons from $\ke3$ decays.  A small residual energy scale mismatch between
data and Monte Carlo is removed by adjusting the energy scale in data
such that the sharp edge in the $z$-vertex distribution at the regenerator
matches between data and Monte Carlo as shown in Fig. \ref{fig:edgematch}.  
The final energy scale adjustment for 1997 data is
shown as a function of kaon energy in Fig. \ref{fig:scalechange};  the
average size of the correction is $\sim$0.04\%.
As a result of improvements to the simulation and reconstruction of clusters, 
the required energy scale adjustment is smaller and less
dependent on kaon energy than in the previous analysis.

\looseness+1
Background to the $\Kpp$ signal modes is simulated using the Monte Carlo, 
normalized to data outside the signal region, and subtracted.  
In this analysis, we use decays 
from coherently regenerated kaons only; any kaons that scattter with non-zero
angle in the regenerator are treated as background.  Scattering background is
the same for both $\Kpm$ and $\Kzz$ decays so it can be identified 
using the reconstructed transverse momentum of the decay products in charged mode;
we use $\Kpm$ decays to tune the simulation of scattering background on which we
must rely in neutral mode. Non-$\pi\pi$ background
is present due to the misidentification of high branching-ratio decay modes
such as $\ke3$, $\km3$, and $\Kzzz$.    
Background contributes less than 0.1\% of
$\kchrg$ data and about 1\% of $\kneut$ data.
Tables \ref{tb:chrgbgfrac} and \ref{tb:neutbgfrac} contain a summary of all the background 
fractions for each year.  There are some variations in background levels among the years due
to differences in trigger and veto requirements.

\begin{table}[ht]
\centering
\begin{tabular}{l|cc|cc}\hline\hline
          & \multicolumn{2}{c|}{Vacuum Beam} & \multicolumn{2}{c}{Regenerator Beam} \\
Source                           & 1997    & 1999    & 1997    & 1999    \\ \hline \hline
Regenerator Scattering           & ---     & ---     & 0.073\% & 0.075\% \\
Collimator Scattering            & 0.009\% & 0.008\% & 0.009\% & 0.008\% \\
$\ke3$                           & 0.032\% & 0.032\% & 0.001\% & 0.001\% \\
$\km3$                           & 0.034\% & 0.030\% & 0.001\% & 0.001\% \\ \hline
Total Background                 & 0.074\% & 0.070\% & 0.083\% & 0.085\% \\ \hline\hline
\end{tabular}
\caption{Summary of $\kchrg$ background levels}
\label{tb:chrgbgfrac}
\end{table}

\begin{table*}[ht]
\centering
\begin{tabular}{l|ccc|ccc}\hline\hline
          & \multicolumn{3}{c|}{Vacuum Beam} & \multicolumn{3}{c}{Regenerator Beam} \\
Source                             & 1996    & 1997    & 1999    & 1996   & 1997    & 1999    \\ \hline \hline
Regenerator Scattering & 0.288\% & 0.260\% & 0.258\% & 1.107\%& 1.092\% & 1.081\% \\
Collimator Scattering  & 0.102\% & 0.122\% & 0.120\% & 0.081\%& 0.093\% & 0.091\% \\
$\Kzzz$                & 0.444\% & 0.220\% & 0.301\% & 0.015\%& 0.006\% & 0.012\% \\
Photon Mispairing      & 0.007\% & 0.007\% & 0.008\% & 0.007\%& 0.008\% & 0.007\% \\
Hadronic Production    & 0.002\% & 0.001\% & ---     & 0.007\%& 0.007\% & 0.007\% \\ \hline
Total Background       & 0.835\% & 0.603\% & 0.678\% & 1.209\%& 1.197\% & 1.190\% \\ \hline\hline
\end{tabular}
\caption{Summary of $\kneut$ background levels.  Note that
photon mispairing is not subtracted from the data and is not included in the total background sum.}
\label{tb:neutbgfrac}
\end{table*}

After all event selection requirements are applied and  
background is subtracted,  
we have a total of 25 million vacuum beam $\kchrg$ decays and 
6 million vacuum beam $\kneut$ decays The numbers of 
events collected in each beam are summarized in Table \ref{tb:eventtot}.

\begin{table}[ht]
\centering
\begin{tabular}{l|cc}
\hline\hline
             & Vacuum Beam & Regenerator Beam \\ \hline
$\kchrg$     & 25107242    & 43674208         \\
$\kneut$     & 5968198     & 10180175         \\ \hline\hline
\end{tabular}
\caption {Summary of event totals after all selection criteria and background
subtraction.}
\label{tb:eventtot}
\end{table}

Table \ref{tb:systsummary} summarizes the systematic uncertainties on
$\reepoe$.  We describe the procedure for evaluating several important 
systematic uncertainties below.

\begin{table}[ht]
\centering
\begin{tabular}{l|cc} 
\hline\hline
Source                & \multicolumn{2}{c}{Error on $\reepoe$ ($\eu$)} \\
                      & $\kchrg$ & $\kneut$                             \\ \hline
Trigger               & 0.23     & 0.20                                 \\
CsI cluster reconstruction & --- & 0.75                                 \\
Track reconstruction  & 0.22     & ---                                  \\
Selection efficiency  & 0.23     & 0.34                                 \\
Apertures             & 0.30     & 0.48                                 \\
Acceptance            & 0.57     & 0.48                                 \\
Background           & 0.20     & 1.07                                 \\
MC statistics         & 0.20     & 0.25                                 \\ \hline
Total                 & 0.81     & 1.55                                 \\ \hline
Fitting               & \multicolumn{2}{c}{0.31}                       \\ \hline
Total                 & \multicolumn{2}{c}{1.78}                          \\ \hline\hline
\end{tabular}
\caption{Summary of systematic uncertainties in $\reepoe$.}  
\label{tb:systsummary}
\end{table}

\emph{Acceptance:}
We use the Monte Carlo simulation to estimate
the acceptance of the detector in
momentum and $z$-vertex bins in each beam.  We evaluate the quality of
this simulation by comparing energy-reweighted $z$-vertex distributions
in the vacuum beam between data and Monte Carlo. We fit a line to the 
data-MC ratio of the $z$-vertex distributions and
call the slope of this line the acceptance ``z-slope.'' 
A z-slope affects the value of $\reepoe$ by producing a bias between the 
regenerator and vacuum beams because of the very different $z$-vertex 
distributions in the two beams; we use the known difference of the mean
$z$ values for the vacuum and regenerator beams along with the measured
z-slope to evaluate the systematic error on $\reepoe$.

\begin{figure}
\begin{center}
\includegraphics[width=80mm]{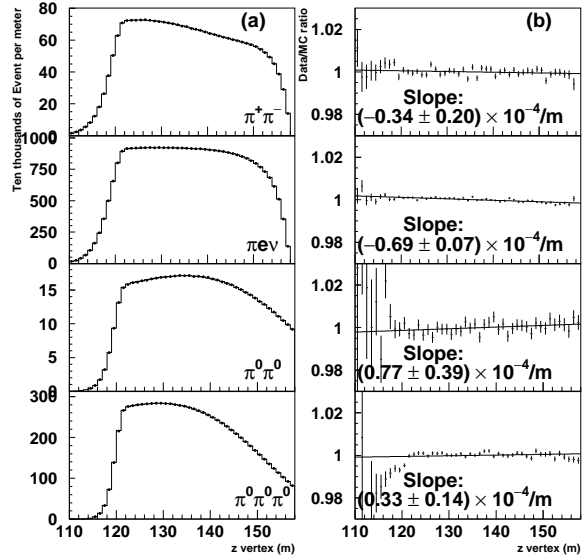}
\caption{Comparison of the vacuum beam $z$ distributions for
data (dots) and MC (histogram).  The data-to-MC ratios on
the right are fit to a line, and the z-slopes (see text) are
shown.  All distributions are for the full data sample used in this
analysis.}
\label{fig:zslopes}
\end{center}
\end{figure}

Figure \ref{fig:zslopes} shows the measured z-slopes for the full $\KLpm$,
$\ke3$, $\KLzz$, and $\Kzzz$ event samples.  We use the $\ppc$ z-slope
to set the systematic uncertainty and measure the $\keth$ z-slope as a
crosscheck.  For neutral mode, we use the high statistics $\zzz$ mode
to set the systematic uncertainty because it has the same type of particles 
in the 
final state as $\pzpz$ and is more sensitive than $\pzpz$ to potential 
problems in the reconstruction due to close clusters, energy leakage at 
the CsI edges, and low photon energies.  

\begin{figure}
\begin{center}
\includegraphics[width=80mm]{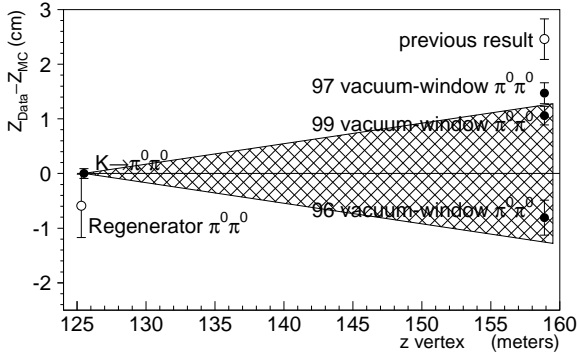}
\caption{Energy scale tests at the regenerator 
and vacuum window.  The difference between the reconstructed $z$ positions 
for data and MC is plotted for the $\kneut$, regenerator $\pzpz$, and vacuum 
window $\pzpz$ samples.  The solid point at the regenerator edge is the 
$\kneut$ sample; there is no difference between data and MC by
construction.  The open point at the regenerator edge is the average shift 
of the regenerator $\pzpz$ samples for all three years.  The points at the 
vacuum window are the shifts for the vacuum window samples for each year 
separately.  The hatched region shows the range of data-MC shifts covered 
by the total systematic uncertainty from the energy scale.
For reference, the data-MC shift at the vacuum window from the 2003
analysis is also plotted.}
\label{fig:escalesyst}
\end{center}
\end{figure}

\emph{Energy Scale:}
The final energy scale adjustment ensures that the energy scale matches 
between data and MC at the regenerator edge, but we must check whether 
the data and MC energy scales remain matched for the full length of the 
decay volume.  
We check the energy scale at the downstream end of the decay region by 
studying the $z$-vertex distribution of $\pzpz$ pairs produced
by hadronic interactions in the vacuum window in data and MC.  
To verify that this type of 
production has a comparable energy scale to $\kneut$, we also study the 
$z$-vertex distribution of hadronic $\pzpz$ pairs produced in the regenerator.
The data-MC comparisons of reconstructed $z$
vertex for these samples are shown in Fig. \ref{fig:escalesyst}.

To convert these shifts to an uncertainty in
$\reepoe$, we consider a
linearly varying energy scale distortion such that no adjustment 
is made at the regenerator edge and the $z$ shift at the vacuum window is 
that measured by the hadronic vacuum window sample.  The average energy scale distortion
we apply is shown by the hatched region in Fig. \ref{fig:escalesyst}.
We rule out energy scale distortions that vary non-linearly as a function 
of $z$ vertex as they introduce data-MC discrepancies in other distributions.
The systematic error on $\reepoe$ due to uncertainties in the $\kneut$ 
energy scale is 0.65$\eu$; this is a factor of two smaller than in the previous
analysis.

\begin{figure}
\begin{center}
\includegraphics[width=80mm]{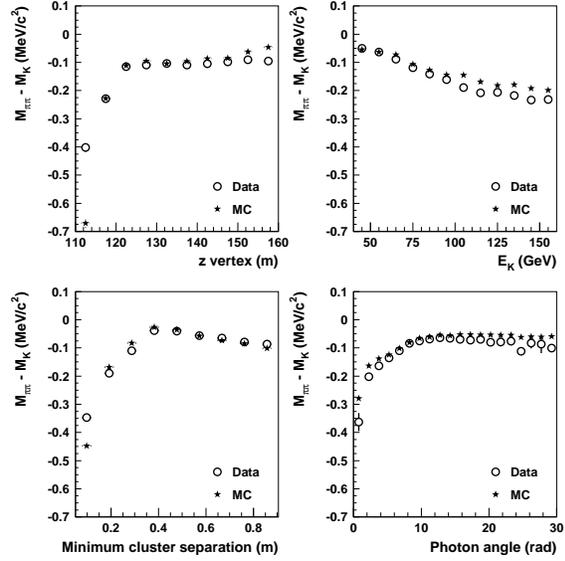}
\caption{Comparisons 
of the reconstructed kaon mass vs $z$-vertex (top left), kaon energy 
(top right), minimum cluster 
separation (bottom left), and photon 
angle (bottom right) for 1999 data and MC.
The values plotted are the difference between the reconstructed
kaon mass for each bin and the nominal PDG kaon mass.}
\label{fig:nonlin1}
\end{center}
\end{figure}

\begin{figure}
\begin{center}
\includegraphics[width=80mm]{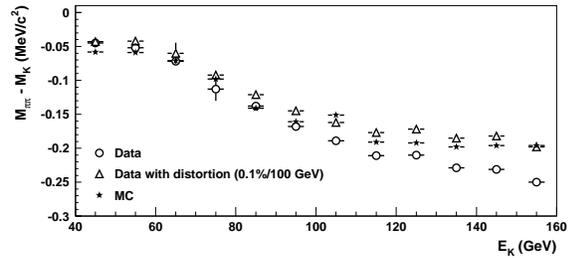}
\caption{Effect of 0.1\%/100 GeV distortion on M$_K$ vs E$_K$ for 1999 data.
The values plotted are the difference between the reconstructed
kaon mass for each bin and the nominal PDG kaon mass.}
\label{fig:nonlin2}
\end{center}
\end{figure}

\emph{Energy Non-linearity:}
Some reconstructed quantities in the analysis do not depend on the CsI
energy scale, but are sensitive to energy non-linearities.
To evaluate the effect of energy non-linearities on the reconstruction,
we study the way the reconstructed kaon mass varies with reconstructed 
kaon energy, 
kaon $z$ vertex, minimum cluster separation, and incident photon angle.  
Data-MC comparisons for these distributions for the 1999 data sample 
are shown in Fig. 
\ref{fig:nonlin1}.  To measure any bias resulting from the nonlinearities
that cause the small data-MC
differences seen in these distributions, we investigate adjustments to the 
cluster energies 
that improve
the agreement between data and MC in the plot of reconstructed kaon mass 
vs kaon energy.  We find that a 0.1\%/100 GeV distortion produces the best 
data-MC 
agreement for the 1997 and 1999 datasets.  
Figure \ref{fig:nonlin2} shows the 
improvement in data-MC agreement with this distortion applied 
to 1999 data.
The data-MC agreement in the reconstructed kaon mass as a function of
kaon energy has been significantly improved compared to the previous analysis 
in which a 0.7\%/100 GeV distortion was required for 1997 data.

\section{Results}
The value of $\reepoe$ and other kaon parameters
$\delm$, $\tauS$, $\phiep$, and $\imepoe$ are determined
using a fitting program. The fitting procedure
is to minimize $\chi^2$ between background subtracted data
yield and a prediction function. The prediction
function uses the detector acceptance determined by the Monte Carlo
simulation. The fits are  performed in $10$~GeV/$c$ kaon momentum
bins. To determine $\reepoe$, we use the $z$ integrated data yield;
to measure the other kaon parameters a $z$-binned fit is performed.

In the fit for $\reepoe$, the inputs are the observed number of 
$\kchrg$ and $\kneut$ decays in each of twelve 10 GeV/c momentum bins.  
The kaon fluxes for
$\kchrg$ and $\kneut$ in each momentum bin, the regeneration parameters,
and $\reepoe$ are free parameters.  CPT symmetry is assumed by setting the
phases $\phipm$ and $\phizz$ equal to the superweak phase.
The final KTeV measurement of $\reepoe$ for the full 1996, 1997,
and 1999 combined dataset is:
\bqa
\reepoe & = & [19.2 \pm 1.1(stat) \pm 1.8(syst)]\eu  \nonumber \\
        & = & [19.2 \pm 2.1]\eu.
\eqa
The fit quality, quantified by 
$\chi^2/\nu = 22.9/21$, is good.

\begin{figure}
\begin{center}
\includegraphics[width=80mm]{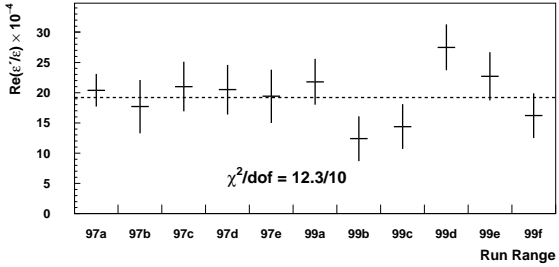}
\caption{$\reepoe$ in subsets of the data sample.  Each point is
statistically independent.  The dashed line indicates the value of $\reepoe$ for the full data sample.  The 97a 
run range includes the 1996 $\kneut$ data.}
\label{fig:eperuns}
\end{center}
\end{figure}

\begin{figure}
\begin{center}
\includegraphics[width=80mm]{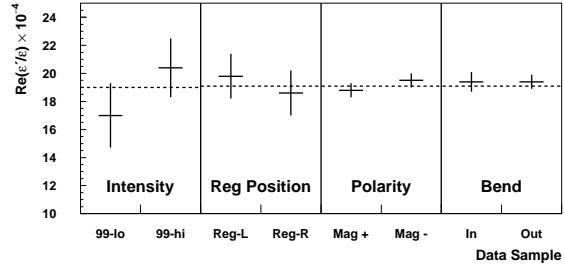}
\caption{$\reepoe$ consistency with beam intensity, regenerator position, 
magnet polarity, and track bend.   The low and high intensity samples are from 1999
only and have
average rates  of $\sim1\times10^{11}$ protons/s and  $\sim1.6\times10^{11}$ 
protons/s, respectively.  
Reg-left and reg-right refer to the position of the regenerator beam in 
the detector.  Mag+ and Mag- are
the magnet polarity and in/out are the bend of the tracks in the magnet.  
In the polarity and bend subsets the $\kneut$ sample is common to both fits; 
the errors are estimated by taking the quadrature difference with
the error for the full dataset.  The dashed lines
indicate the value of $\reepoe$ in the appropriate full data sample.}
\label{fig:epechecks}
\end{center}
\end{figure}

\begin{figure}
\begin{center}
\includegraphics[width=80mm]{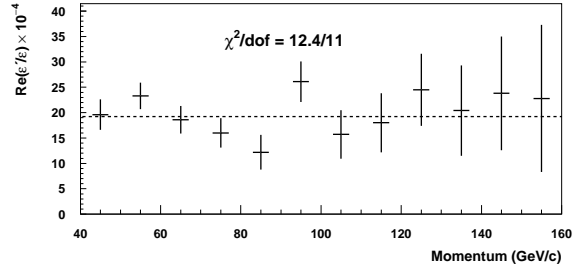}
\caption{$\reepoe$ in 10 GeV/c momentum bins.
The dashed line indicates the value for the full data sample.}
\label{fig:epepbins}
\end{center}
\end{figure}

We perform several checks of our result by breaking the data 
into subsets and checking the
consistency of the $\reepoe$ result.  To check for 
any time dependence, we break the data into 11 run ranges with roughly equal 
statistics.  We divide the data based on beam intensity, regenerator
position, magnet polarity, and direction in which the tracks bend in the
magnet.  We check for dependence of the result on kaon momentum by breaking 
the data into twelve 10 GeV/c momentum bins.
The $\reepoe$ 
results for these tests are shown in Figs. \ref{fig:eperuns}, 
\ref{fig:epechecks}, and \ref{fig:epepbins}.  We find 
consistent results in all of these subsamples.

The combined result for 1996 and 1997 data only is 
$\reepoe = [20.0 \pm 1.7(stat)] \eu$.
This result is consistent with the previously published KTeV result, which
is based on the same subset of data,
of $\reepoe =  [20.7 \pm 1.5(stat)] \eu$ \cite{prd03}.

\begin{figure}
\begin{center}
\includegraphics[width=80mm]{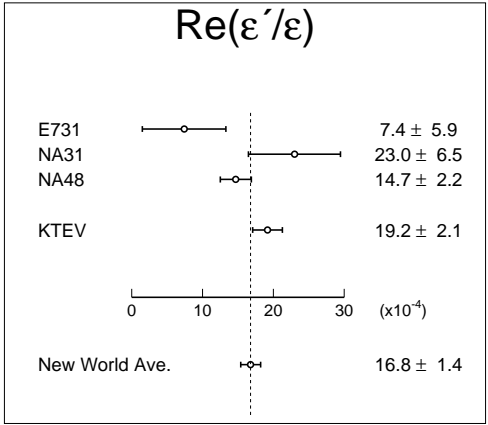}
\caption{New world average for $\reepoe$ combining results from
E731\cite{prl:731}, NA31\cite{pl:na31}, NA48\cite{na48:reepoe},
and KTeV.}
\label{fig:exptcomp}
\end{center}
\end{figure}

The value of $\reepoe$ is consistent with other experimental results.  
The weighted 
average of the new KTeV result with previous measurements is
$\reepoe  =  [16.8 \pm 1.4]\eu$; see Fig. \ref{fig:exptcomp}.  The 
consistency probability of these results is $13\%$.

The regenerator beam decay distribution is sensitive to the kaon parameters 
$\tauS$, $\delm$,
$\phiep$, $\reepoe$, and $\imepoe$.
We measure these parameters by fitting the decay vertex distribution in 
the regenerator beam using a single, z-binned fit.  
The five kaon parameters, $\tau_S$, $\delm$, $\phiep$, 
$Re(\epsilon'/\epsilon)$, and $Im(\epsilon'/\epsilon)$ are free
parameters of the $z$-binned fit.  The fit thus provides the most
general description of the data with no requirement of CPT invariance.
All the systematic uncertainties are evaluated for the fit using
a procedure identical to that used for the $\reepoe$ measurement, 
and accounting 
for correlations between the parameters. CPT invariance is imposed
{\it a posteriori}, including the total errors of the parameters with their
correlations, to obtain a precise measurement of $\delm$ and $\tau_S$.

This approach allows a self-consistent analysis of the data with and without
CPT constraints. The results are crosschecked with
separate fits for $\delm$ and $\tau_S$ 
performed  with CPT invariance  imposed {\it a priori}.  The two procedures
agree to within $1.3 \sigma_{\rm stat}$ and the total uncertainties
agree to within $\sim 10\%$.

\begin{figure}
\centering
\includegraphics[width=80mm]{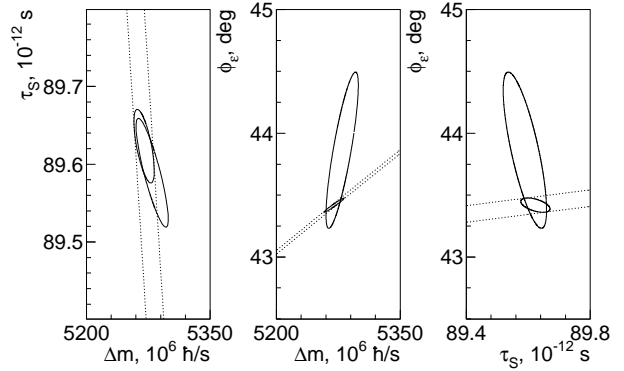}
\caption{$\Delta \chi^2 = 1$ contours of total uncertaitny
for (a) $\delm$-$\tauS$, (b) $\phiep$-$\delm$ and (c) $\tauS$-$\phiep$.
Larger  ellipses correspond to the $z$-binned fit without  CPT invariance assumption.
Dashed lines correspond to $\phiep = \phi_{SW}$ CPT constraint. Smaller
ellipes are obtained after applying this constraint.}
\label{fig:swdmts_full}
\end{figure}

\begin{figure}
\centering
\includegraphics[width=80mm]{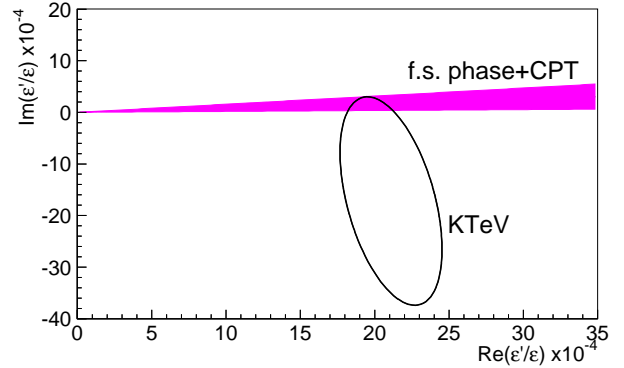}
\caption{$\Delta \chi^2 = 1$ contour
for $Re(\epsilon'/\epsilon)$ vs $Im(\epsilon'/\epsilon)$ as measured by KTeV 
compared to the measurement of $\pi\pi$ phase shifts~\cite{ochs} and the CPT 
invariance expectation.}
\label{fig:reim}
\end{figure}

The results of the single z-binned fit are:
\begin{equation}\label{eq:zresults}
\begin{array}{lcl}
  \delm\,|_{\rm cpt}  &=& [\dmswval \pm \dmswerr ] \times 10^{-12}~{\rm s}, \\
  \tauS\,|_{\rm cpt}  &=& [\tsswval \pm \tsswerr ] \times 10^6 {\rm \hbar/s}, \\
  \phipm &=& [\phpmval \pm \phpmerr]\degs,\\
  \phizz &=& [\phzzval \pm \phzzerr]\degs, \\
  \delta \phi &=& \phiep - \phi_{SW} = [\dphswval \pm \dphswerr]\degs, \\
  \Delta \phi &=& - 3 \imepoe = [\dphcptval \pm \dphcpterr ]\degs.\\
\end{array}
\end{equation}
The fit quality is good: $\chi^2/\nu = 425.4/(432-33)$.
The correlations among these
results are shown in Fig. \ref{fig:swdmts_full} and Fig. 
\ref{fig:reim}.
The value of $\delta \phi = \phiep - \phi_{SW}$ is consistent with 
zero as expected from CPT invariance in $\Kz$-$\Kzbar$ mixing and the
value of $\Delta\phi$ is consistent with zero as expected from CPT
invariance in a decay amplitude.

\section{Conclusion}
Using the full data sample of the KTeV experiment, 
we have made improved
measurements of direct CP violation and other parameters
of the neutral kaon system.
All of these results supersede previous KTeV results.

Assuming CPT invariance, we measure the direct CP violation parameter
\bqa
\reepoe  & = & [19.2 \pm 1.1(stat) \pm 1.8(syst)]\eu \\ \nonumber
        & = & [19.2 \pm 2.1]\eu.  
\eqa
Also under the assumption of CPT invariance, we report new measurements
of the $\KL-\KS$ mass difference and the $\KS$ lifetime:

\begin{equation}
\begin{array}{lcl}
  \delm  &=& [\dmswval \pm \dmswerr ] \times 10^{-12}~{\rm s}\\
  \tauS   &=& [\tsswval \pm \tsswerr ] \times 10^6 {\rm \hbar/s}. \\
\end{array}
\end{equation}

To test CPT symmetry, we measure the phase differences
\begin{equation}
\begin{array}{lcl}
\Delta \phi &=& - 3 \imepoe\\
            &=& [\dphcptval \pm \dphcpterr ]\degs,
\end{array}
\end{equation}
and
\begin{equation}
\begin{array}{lcl}
\phiep - \phi_{SW}            
            &=& [\dphswval \pm \dphswerr]\degs.
\end{array}
\end{equation}
These phase results are consistent with CPT invariance in both the decay 
amplitudes and $\Kz-\Kzbar$ mixing.

After decades of experimental effort, direct CP violation in the neutral
kaon system has now been measured with an uncertainty of about 10\%. 
Considerable improvement in theoretical calculations of $\reepoe$
will be required to take advantage of this experimental precision. 
There is some optimism, however, that the next rounds of calculations 
using lattice gauge theory may approach a 10\%\ uncertainty, making the 
precise measurements of $\epe$ equally precise tests of the Standard Model.

\begin{acknowledgments}
We gratefully acknowledge the support and effort of the Fermilab staff
and the technical staffs of the participating institutions.  This work
was supported in part by the U.S. Department of Energy, The National
Science Foundation, and the Ministry of Education and Science in Japan.

\end{acknowledgments}

\bigskip 

\end{document}